\documentclass[twocolumn,showpacs,preprintnumbers,amsmath,amssymb]{revtex4}
\usepackage{dcolumn}% Align table columns on decimal point
\usepackage{bm}% bold math
\usepackage[dvipdf]{graphicx}% Include figure files
\DeclareGraphicsExtensions{.jpg,.pdf,.mps,.png,.eps,.ps,.EPS}
\begin{document}
\newcommand{\avg}[1]{\langle{#1}\rangle}
\newcommand{\Avg}[1]{\left\langle{#1}\right\rangle}
\def\be{\begin{equation}}
\def\ee{\end{equation}}
\def\bc{\begin{center}} 
\def\ec{\end{center}}
\def\bea{\begin{eqnarray}}
\def\eea{\end{eqnarray}}

\title{Emergence of large cliques in random scale-free networks}
\author{Ginestra Bianconi$^{1}$ and Matteo Marsili$^1$}
\affiliation{$^1$The Abdus Salam International Center for Theoretical Physics, Strada Costiera 11, 34014 Trieste, Italy } 
\begin{abstract}
In a network cliques are fully connected subgraphs that  reveal which are the  tight communities present in it. Cliques of size  $c>3$ are present in random Erd\"os and Renyi graphs only in the limit of diverging average connectivity.
Starting from the finding that real scale free graphs have large cliques, we study the clique number in uncorrelated scale-free networks finding both upper and lower bounds. Interesting we   find that in scale-free networks large cliques appear also when the average degree is finite, i.e. even for networks with power-law degree distribution exponents $\gamma \in (2,3)$. Moreover as long as $\gamma<3$ scale-free networks have a maximal clique which diverges with the system size.
\end{abstract}
\maketitle

Scale-free graphs have been recently found to encode the complex structure of many different systems ranging from the Internet to the protein interaction networks of various organisms \cite{RMP,Doro,Internet}.
This topology is clearly well distinguished from the Erd\"os and Renyi (ER) \cite{Janson} random graphs in which every couple of nodes have the same probability $p$ to be linked. In fact while scale-free graphs have a power-law degree distribution $P(k)\sim k^{-\gamma}$ and a diverging second moment $\langle k^2\rangle$ when $\gamma<3$, ER graphs have a Poisson degree distribution and consequently finite fluctuations of the nodes degrees.
The degree distribution strongly affects the statistical properties of processes defined on the graph. For example,
percolation and epidemic spreading which have very different phenomenology when defined on a  ER graph or on a  scale-free graphs \cite{Havlin,Vespi_epi}.

The occurrence of a skewed degree distribution has also striking consequences regarding the frequency of particular subgraphs present in the network.
For example, ER graphs with finite average connectivity have a finite number of finite loops 
\cite{Janson,Monasson}. On the contrary scale-free graphs have a number of finite loops which increases with 
the number $N$ of vertices, provided
that $\gamma\leq 3$ \cite{Loops,lungo}. The abundance of some
subgraphs of small size -- the so-called {\em motifs} -- in biological networks 
has been shown to be related to important functional properties 
selected by evolution \cite{Milo,Vazquez_m,Dobrin}. Among
subgraphs, cliques play an important role. A clique of size $c$ is a
complete subgraph of $c$ nodes, i.e. a subset of $c$ nodes each of
which is linked to any other. The maximal size $c_{\max}$ of a clique in a graph
is called the clique number. Finding the clique number of a generic
network is an NP-complete problem \cite{NPcomplete}, even though it is relatively easy to find upper ($c_+$) and lower ($c_-$) bounds \cite{nota_bounds}. The clique number also 
provides a lower bound for the chromatic number of a graph, i.e. the minimal 
number of colors needed to color the graph \cite{coloring}. Finally, 
cliques and overlapping succession of cliques have been recently used to characterize the community
structure of networks \cite{Vicsek1,Vicsek2}.

\begin{figure}
\includegraphics[width=75mm, height=55mm]{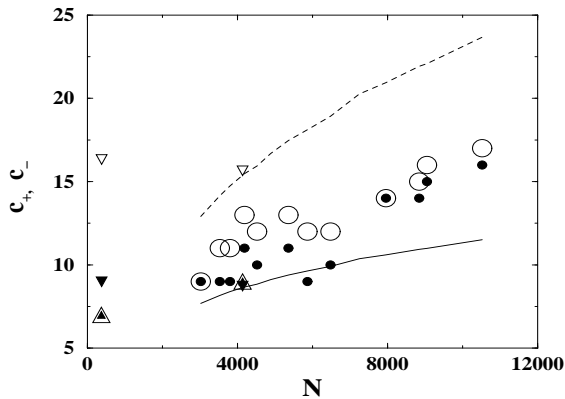}
\caption{
The lower bound $c_{-}$ (filled symbols) and the upper bound  $c_{+}$ (empty symbols) of the clique number of  the Internet graphs(circles) and  the protein interaction networks of e.coli and yeast (triangles) \cite{data_info} are shown as a function of the network size $N$. The  lines (null hypothesis on Internet data) and the triangles pointing down (null hypothesis on protein  interaction networks) indicates the upper bound (dashed line and empty symbols)  and the lower bound (solid line and filled symbols)  computed  from Eq. $(\ref{cbar})$ for random graphs constructed with the same properties of the considered real graphs.}
\label{data} 
\end{figure}

In ER graphs it is very easy to show that cliques of size $3<c\ll N $ appear in the graph only when 
the average degree diverges as $\avg{k}\sim N^{\frac{c-3}{c-1}}$ with $N$ \cite{Janson}. On the other hand, real scale free networks, such as the Internet at the autonomous system level, contain cliques of size much larger than $c=3$. For example, Fig. $\ref{data}$ reports upper and lower bounds $c_{+}, c_{-}$ \cite{nota_bounds} for the  size of the  maximal clique  of the Internet and protein interaction networks of c.elegans and yeast \cite{data_info}. 
This shows that scale-free networks can have large cliques and that the clique number of the Internet graphs increase with the network size $N$. 

Is the presence of such large cliques a peculiar property of how these networks are wired or is this a typical
property of networks with such a broad distribution of degrees? This letter addresses this question and shows that scale free random networks do indeed contain cliques of size much larger than $c=3$. We shall do this by computing the first two moments of the number ${\cal N}_c$ of cliques of size $c$ in a network of $N$ nodes. These provide upper and lower bounds for the probability $P({\cal N}_c>0)$ of finding cliques of size $c$ in a network through the inequalities \cite{Janson}
\begin{equation}\label{bounds}
  \frac{\avg{{\cal N}_c}^2}{\avg{{\cal N}_c^2}}\le P({\cal N}_c>0)\le \avg{{\cal N}_c}.
\end{equation}
Here and in the following the notation $\avg{\ldots}$ will be used for statistical averages.
Eq. (\ref{bounds}) in turn provide upper and lower bounds for the clique number
$\underline{c}\le c_{\max}\le\overline{c}$: Indeed if $\avg{{\cal N}_c}\to 0$ for $c>\overline{c}$ as $N\to\infty$, we can conclude that no clique of size larger than $\overline{c}$ can be found. Likewise if for $c=\underline{c}$ the ratio ${\avg{{\cal N}_c}^2}/\avg{{\cal N}_c^2}$ stays finite, then cliques of size $c\leq\underline{c}$ can be found in the network with at least a finite probability. The results indicate that the finding in Fig. \ref{data} are expected, given the scale free nature of these graphs. Our predictions are summarized in Table \ref{table}. We find that 
the ER result $c_{\rm max}=3$ extends to random scale free networks with $\gamma>3$ whereas for $\gamma<3$ the clique number $c_{\rm max}$ diverges with the network size $N$ in a way which is extremely sensitive of the degree distribution of mostly connected nodes, i.e. to the precise definition of the cutoff. 

The results of Table \ref{table} are derived for the hidden variable ensemble proposed in Ref. \cite{HV1,HV2}, where the link probability $p$ between two nodes is replaced by a function $r(q_i,q_j)$ which depends on the fitness $q_i$ and $q_j$ of the end nodes $i$ and $j$. Apart from its close relation with the ER ensemble, this choice is also convenient because it allows for a simple generalization of the results to networks with a correlated degree distribution  \cite{next}. Quite similar results can be derived for the Molloy-Reed ensemble \cite{MR} with the same approach (provided a cutoff is chosen appropriately to avoid double links among mostly connected nodes) . Other ensembles, such as that of Ref. \cite{Kahng} instead implicitly introduce a degree correlations for highly connected nodes and therefore require a different approach \cite{next}. Given the extreme sensitivity of the clique number on details of the cutoff of the degree distribution, we also expect quite different results.

\begin{table}
\centering
\begin{tabular}{|c|c|c|}
\hline
\hline
  & {$\epsilon=0$} & {$\epsilon\neq 0$}\\
\hline
$\gamma>3$ &\multicolumn{2}{|c|}{$c_{max}=3$}\\ \hline
$2<\gamma<3$ &{$\underline{c}\leq c_{max}\leq \bar{c}$} &{$\underline{c}\leq c_{max}\leq \bar{c}$}\\ 
&  & \\
& {$\bar{c}\simeq \sqrt{b} N^{\frac{3-\gamma}{4}} $} & {$\bar{c}\simeq\frac{3-\gamma}{2}\frac{\log(N)}{|\log(1-\epsilon)|}$} \\ 
  & {$\underline{c}\simeq \alpha \bar{c}^{2/3}$ } & {$\underline{c}=(1-\alpha) \bar{c}$}  \\
 & & \\ \hline
$1<\gamma\leq 2$ & {$\underline{c}\leq c_{max}\leq \bar{c}$} &{$\underline{c}\leq c_{max}\leq \bar{c}$}\\ 
&  & \\
& {$\bar{c}\simeq \sqrt{b'} N^{\frac{1}{2\gamma}}  $} & {$\bar{c}\simeq\frac{1}{\gamma}\frac{\log(N)}{|\log(1-\epsilon)|}$} \\ 
&  & \\
 & {$\underline{c}\simeq \alpha \bar{c}^{2/3}$} & {$\underline{c}= (1-\alpha) \bar{c}$ }  \\  & & \\ \hline
\hline 
\end{tabular}
\caption{Scaling of the theoretically estimated upper and lower bound of the clique number of random scale-free networks with different exponents $\gamma$ of the degree distribution. The precise definitions of $\bar{c}$ and $\underline{c}$ together with the expression for the constants $b, b'$ are given in the text.}\label{table}
\end{table}

{\em Hidden variable network ensemble} 
As in Ref. \cite{HV1} we generate a realization of a scale-free networks by the following procedure: {\em i)} assign to each node $i$ of the graph a hidden continuous variable $q_i$ distributed according a $\rho(q)$ distribution. Then {\em ii)} each pair of nodes with hidden variables $q,q'$ are linked with probability $r(q,q')$. For random scale-free networks with uncorrelated degree distribution, we take $\rho(q)=\rho_0 q^{-\gamma}$ for $q\in [m,Q]$ and 

\begin{equation}\label{rqq}
  r(q,q')=\frac{qq'}{\langle q\rangle N}.
\end{equation}
The average degree $\avg{k}=\avg{q}$ is equal to the average fitness, and it diverges
as $N\to\infty$ for $\gamma<2$. Likewise, the degree $k_i$ of node $i$ follows a Poisson distribution with average $q_i$. Notice that a cutoff is needed in $\rho(q)$ to keep the linking probability $r(q,q')$ smaller than one. In particular, we will take require 
\begin{equation}\label{cutoff}
  Q=(1-\epsilon)\sqrt{\langle q\rangle N}
\end{equation}
so that $r(Q,Q)=1-\epsilon$.
For $\gamma>3$, values of $q_i\approx Q$ will never occur, as the maximal $q_i\approx N^{1/(\gamma-1)}\ll Q$. We shall see that this is immaterial for the clique number, however. Instead, for $\gamma<2$, $\avg{q}$ diverges with the cutoff, and hence $Q\sim N^{1/\gamma}$.

{\em Average number of cliques.}
A clique of size $c$ is a set of $c$ distinct nodes ${\cal C}=\{i_1,\ldots,i_c\}$, each one connected with all the others. For each choice of the nodes, the probability that they are connected in a clique is
\be
\prod_{i\neq j\in{\cal C}} r(q_{i},q_{j})=\prod_{i\in{\cal C}}\left(\frac{q_i}{\sqrt{\avg{q}N}}\right)^{c-1}
\ee
where we used Eq. (\ref{rqq}). Fixing a small fitness interval $\Delta q$, let $n(q)$ be the number of nodes $i\in {\cal C}$ with fitness $q_i\in (q,q+\Delta q)$. The number of ways in which we can pick $c$ nodes in the network with $n(q)$ nodes with fitness $q$ can be expressed by combinatorial factors. Hence, with the shorthand ${\cal Q}=q/\sqrt{\langle q \rangle N}$,
\bea
\langle{\cal N}_c \rangle &=&\sum_{\{n(q)\}}' \prod_{q} \left(\begin{array}{l} N(q)\\ n(q) \end{array}
\right) {\cal Q}^{(c-1)n(q)}
\eea
where the sum is extended to all the sequences  $\{n(q)\}$  satisfying $\sum_q n(q)=c$. 
Introducing such constraint by a delta function, we can perform the resulting integral by saddle point
method, i.e.
\bea
\langle{\cal N}_c \rangle = \int_{-\pi}^\pi \frac{d\omega}{2\pi}e^{Nf(i\omega)}\simeq
\frac{e^{Nf(y^*)}}{\sqrt{2\pi N|f''(y^*)|}}
\label{N_m}
\eea
where $f(y)=\frac{c}{N}y+\Avg{\log \left[1+{\cal Q}^{c-1} e^{-y}\right]}$, 
and we have taken the  limit $\Delta q\rightarrow 0$. In Eq. (\ref{N_m}) $y^*$ is fixed by the saddle point condition 
\be
\frac{c}{N}=\left\langle\frac{{\cal Q}^{c-1} e^{-y^*}}{1+{\cal Q}^{c-1} e^{-y^*}}\right\rangle.
\label{ystar}
\ee
We present here an asymptotic estimate of $\avg{{\cal N}_c}$. Slightly more refined arguments, which do not add much
to the understanding given here, can be used to derive an upper bound \cite{next}.
In the limit $N\to\infty$, the left hand side of Eq. (\ref{ystar}) is small, hence to a good approximation 
$c\approx N\avg{{\cal Q}^{c-1}}e^{-y^*}$ \cite{nota_exp}. Inserting this in Eq. (\ref{N_m}) we find

\be
\Avg{{\cal N}_c}\approx \left(\frac{Ne\avg{{\cal Q}^{c-1}}}{c}\right)^c \sqrt{\frac{2\pi}{c}}.
\label{Nml}
\ee
Therefore, in order to have $\avg{{\cal N}_c}\to 0$ it is sufficient to take $c>\bar{c}$, where $\bar{c}$ is the solution of 
\be
Ne\avg{{\cal Q}^{c-1}}=c.
\label{cbar}
\ee
We consider now separately the case of scale-free networks with different exponents $\gamma$ of the degree distribution.
\begin{itemize}
\item{Networks with $\gamma>3$}\\
Eq. (\ref{cbar}) has no solution for $c>\gamma$. Indeed $N\avg{{\cal Q}^{c-1}}\sim N^{(3-\gamma)/2}\to 0$ in this range. For $c<\gamma$, the integral in $\avg{{\cal Q}^{c-1}}$ is no longer dominated by the upper cutoff, and it is hence finite. Therefore $N\avg{{\cal Q}^{c-1}}\sim N^{(3-c)/2}$ which implies that $\bar{c}=3$. It is easy to see that this conclusion holds also if we take the natural cutoff $Q=a N^{\frac{1}{\gamma-1}}$.

\item{Network with $2<\gamma<3$}\\
Using Eq. (\ref{cutoff}), Eq. (\ref{cbar}) becomes

\begin{equation}\label{cbar23}
  \frac{\bar{c}(\bar{c}-\gamma)}{(1-\epsilon)^{\bar{c}-\gamma}}\simeq bN^{(3-\gamma)/2}
\end{equation}
for $b= (\gamma-1)m^{(\gamma-1)}e \avg{q}^{(1-\gamma)/2}$. The solution depends crucially on whether $\epsilon=0$ or not. In the former case $\bar{c}\sim N^{(3-\gamma)/4}$ increases as a power law of the system size, whereas for $\epsilon>0$ it increases only as $\log N/\log(1-\epsilon)$,
as detailed in Table \ref{table}.

\item {Network with $1<\gamma<2$}\\
Taking into account the divergence of $\avg{q}$ and $Q\sim N^{1/\gamma}$, Eq. (\ref{cbar}) becomes

\begin{equation}\label{cbar12}
  \frac{\bar{c}(\bar{c}-\gamma)}{(1-\epsilon)^{\bar{c}-\gamma}}\simeq b'N^{1/\gamma}
\end{equation}
with $b'=\{(\gamma-1)[m(2-\gamma)]^{(\gamma-1)}\}^{1/\gamma}$ . Again, for $\epsilon=0$ and $\epsilon>0$ we find different results, $\bar{c}\sim N^{1/(2\gamma)}$ 
and $\bar{c}\sim \log N/\log(1-\epsilon)$ respectively (see Table \ref{table}).
\end{itemize}

{\em Second moment of the average number of cliques.}
When computing the average number of some particular subgraphs in a random network ensemble the result might be dominated be extremely rare graphs with an anomalously large number of such subgraphs. In this cases, the average  number of a  subgraph does not provide a reliable indication of its value.
In order to have more insight on the characteristics of typical networks we use the classical relation Eq. (\ref{bounds}) of probability theory \cite{Janson} which provides a lower bound for the probability that a typical graph contains at least one clique of size $c$. This requires us to compute the second moment $\langle {\cal N}^2_c\rangle $ of the number of cliques of size $c$ in the random graph ensemble. 
In order to do this calculation we are going to count the average number of pairs of cliques of size $c$  present in the graph with an overlap of $o=0,\dots, c$ nodes. We use the notation $\{n(q)\}$ to indicate the number of the nodes with fitness $q$ belonging to the first  clique, $\{n_o(q)\}$ to indicate the number of nodes belonging to the overlap and  $\{n'(q)\}$ to indicate  the number of nodes belonging to the second clique but not to the overlap. We consider only sequences $\{n(q)\}, \{n'(q)\}, \{n_o(q)\} $ which satisfy $\sum_q n(q)=c$, $\sum_q n_o(q)=o$ and  $\sum_q n'(q)=c-o$.
With these conditions, following the same steps as for $\langle {\cal N}_c\rangle$ we get
\bea
\langle{\cal N}^2_c \rangle &=&\sum_{o=0}^c\int dy \int dy^{o} \int dy' e^{N\langle f\left(y,y',y^o,{\cal Q}\right)\rangle}
\eea
where 
\bea
&f(y, y',y^o,{\cal Q})=\frac{1}{N}[y c+y'(c-o)+y^{o}o]+\nonumber\\
&+\log \left[1+\left(e^{-y'}+e^{-y}\right){\cal Q}^{c-1}+ e^{-(y+y^o)}{\cal Q}^{2c-o-1} \right].
\eea

The evaluation of this integral by saddle point is  straightforward. The key idea is that, in order to have $\avg{{\cal N}_c^2}$ of the same order as $\avg{{\cal N}_c}^2$ one needs to require that the sum is dominated by configurations with non-overlapping cliques ($o\sim 0$). Using the estimate of $\avg{{\cal N}_c}$ derived above and the definition of $\bar{c}$, for $\gamma<3$ we arrive at 
\bea
P({\cal N}_{\hat {c}}>0)\geq \frac{{\langle{\cal N}_c \rangle}^2}{{\langle{\cal N}^2_c \rangle}}&\geq \left[1+ \frac{c(c-\gamma) (1-\epsilon)^{(\bar{c}-c)}e}{\bar{c}(\bar{c}-\gamma)}\right]^{-c}.
\label{n2_lowerb.eq}
\eea

The lower bound for the clique number will depend on $\epsilon$ and $\bar{c}$.

In the case  $\epsilon=0$ 
lets  define the clique size $\underline{c}$ satisfying 
\be
\frac{\underline{c}(\underline{c}-\gamma)e}{\bar{c}(\bar{c}-\gamma)}=\frac{1}{\underline{c}}
\label{clow}
\ee
i.e. $\underline{c}\sim \bar{c}^{2/3}$.
From Eq.  $(\ref{n2_lowerb.eq})$ and the definition of $\underline{c}$  it follows that as $N,\bar{c}\rightarrow \infty$ the probability to
 have at least a clique  of size $c= \underline{c}$ is finite, i.e.
\bea
P({\cal N}_{\underline {c}}>0)\geq \frac{1}{e}. 
\eea
Instead in  the case  $\epsilon>0$ for any $\alpha>0$ the r.h.s. of Eq. ($\ref{n2_lowerb.eq}$) is very close to $1$
 for and   clique sizes $\underline{c}= (1-\alpha)\bar{c}$ and $\bar{c}\gg 1/(\alpha \epsilon)$,i.e.
\bea
P({\cal N}_{\underline{c}}>0) \rightarrow 1.
\eea
This implies that for $\epsilon>0$ the lower bound is very close to the upper bound $\underline{c}=(1-\alpha)\bar{c}$
for very large networks.

{\em Conclusions} In conclusion we have calculated upper and lower
bounds for the maximal clique size $c_{max}$ in uncorrelated
scale-free network, showing that $c_{\max}$ diverges with the network
size $N$ as long as $\gamma<3$. In particular large cliques are
present in scale-free networks with $\gamma\in (2,3)$ and finite
average degree. It is suggestive to put the emergence of large cliques
for $\gamma<3$ in relation with the persistence up to zero temperature
of long range order in spin models defined on these graphs
\cite{Ising_model_on_graph}. These results were derived within the
hidden variable ensemble \cite{HV1,HV2}, but the same method can be
extended to other ensembles \cite{MR,Kahng} including those with a
correlated degrees. 

In Fig. \ref{data} we compare the upper and lower bounds derived here
for random scale-free graphs with the estimated clique number of real
networks. These networks have many nodes with degree larger than that
of the structural cutoff.  Networks with such highly connected nodes
cannot be considered as uncorrelated. The best approximation, within
the class of uncorrelated networks discussed here, is provided by
those with maximal cutoff ($\epsilon=0$). The bounds of Fig.
\ref{data} have been derived from Eq. (\ref{cbar}) and (\ref{clow}),
assuming a random network with {\em i)} an exponent $\gamma$ as
measured from real data {\em ii)} the same number of nodes and links
(i.e. the same average degree) and {\em iii)} a structural cutoff
given by Eq. (\ref{cutoff}) with $\epsilon=0$. Also notice that
$\epsilon=0$ yields the least stringent bounds. 

Fig. \ref{data} shows that generally the largest clique size
$c_{\max}$ of real networks falls well within our bounds.  Of course,
accounting for the presence of correlations in the degree of highly
connected nodes in these networks may provide more precise estimates.
We saw that our estimates are very sensitive to the tails of
the degree distribution and we expect it to depend also strongly on
the nature of degree correlations. Preliminary results,
extending the present calculation to correlated networks \cite{Kahng}
where $r(q,q')=1-e^{-\alpha q q'}$ with the natural cutoff
$Q\simeq N^{1/(\gamma-1)}$, indicates that the clique number
can take values a factor two bigger than in real data \cite{next}.
These preliminary results underline the importance of extending this
approach to correlated networks.

G. B.  was partially supported by EVERGROW and by EU grant HPRN-CT-2002-00319, STIPCO.


\begin{thebibliography}{}
\bibitem{nota_bounds} Let $K$ be maximal integer such that removing iteratively all nodes with degree less than $K$ leaves a non-empty sub-graph, the so-called $K$-core, and let $c_K$ be the clique number of the $K$-core. Then if $c_K>K$ then it is easy to show that $c_{\max}=c_K$. Else if $c_K\le K$ then we have $c_K\le c_{\max}\le K$. Hence we derive the bounds $c_-=\min(c_K,K)\le c_{\max}\le\max(c_K,K)$.
\bibitem{data_info}
The Internet datasets  we used  are the ones collected by University of Oregon Route Views project, NLANR and the protein-protein interaction datasets are one listed in DIP database.
\bibitem{nota_exp} Indeed Eq. (\ref{ystar}) can be expanded in moments of ${\cal Q}^{c-1}$ and a direct calculation shows that the
ratio of the $n^{\rm th}$ term to the first is $\avg{{\cal Q}^{(c-1)n}}e^{-ny^*}/\avg{{\cal Q}^{c-1}}e^{-y^*}\approx\frac{c-\gamma}{(c-1)n-\gamma+1}\left(\frac{c(c-1)}{b N}\right)^{n-1}$, which vanishes as $N\to\infty$ when $c\approx \bar{c}$.
\end{thebibliography}
\end{document}